\documentclass[twocolumn,secnumarabic,amssymb,aps,prl,longbibliography,nofootinbib,superscriptaddress]{revtex4-2}

\usepackage{graphicx}
\usepackage{amsmath}
\usepackage{subfigure}
\usepackage{color}
\usepackage{bm}

\allowdisplaybreaks[4]

\usepackage[colorlinks=true,linkcolor=blue,citecolor=blue, urlcolor=blue]{hyperref}

\begin{document}

\title{Renormalization and Factorization Scale-Invariant Predictions for the Higgs Rare Decay $H\to J/\psi+\gamma$ via the Principle of Maximum Conformality}

\author{Qi-Sha Ran}
\email{ranqs@stu.cqu.edu.cn}
\affiliation{Department of Physics, Chongqing Key Laboratory for Strongly Coupled Physics, Chongqing University, Chongqing 401331, P.R. China}

\author{Xing-Gang Wu}
\email{wuxg@cqu.edu.cn}
\affiliation{Department of Physics, Chongqing Key Laboratory for Strongly Coupled Physics, Chongqing University, Chongqing 401331, P.R. China}

\author{Jiang Yan}
\email{yjiang@itp.ac.cn}
\affiliation{Institute of Theoretical Physics, Chinese Academy of Sciences, Beijing, 100049, P.R. China}

\author{Xu-Chang Zheng}
\email{zhengxc@cqu.edu.cn}
\affiliation{Department of Physics, Chongqing Key Laboratory for Strongly Coupled Physics, Chongqing University, Chongqing 401331, P.R. China}

\author{Chang-Xin Liu}
\email{liucx@cqu.edu.cn}
\affiliation{Department of Physics, Chongqing Key Laboratory for Strongly Coupled Physics, Chongqing University, Chongqing 401331, P.R. China}

\date{\today}

\begin{abstract}

We investigate the \(J/\psi\) direct production mechanism in the rare exclusive Higgs decay \(H\to J/\psi+\gamma\) within nonrelativistic QCD (NRQCD), which provides a clean probe for extracting the charm-quark Yukawa coupling to the Higgs boson. The Principle of Maximum Conformality (PMC) is used to remove conventional renormalization-scheme and scale ambiguities in the next-to-next-to-leading-order (N\(^2\)LO) perturbative QCD series. Large logarithmic contributions arising from Yukawa coupling renormalization are resummed, providing a reliable foundation for subsequent analyses. Using the experimentally measured leptonic decay width of \(J/\psi\) and the N\(^2\)LO perturbative result, we extract the factorization-scale-dependent long-distance matrix element \(\langle J/\psi({\bm \epsilon})|\psi^{\dagger}{\bm \sigma}\cdot{\bm \epsilon}\chi(\mu_\Lambda) |0\rangle\). Combining this with the factorization-scale-dependent short-distance coefficient, we obtain a factorization-scale-invariant decay width for the channel. Compared with earlier predictions in the literature, our fixed-order result for \(\Gamma(H\to J/\psi+\gamma)\) is more robust and precise, with good convergence and no renormalization- or factorization-scale dependence. We find \(\Gamma(H\to J/\psi+\gamma) = (6.4574^{+0.3995}_{-0.3995}) \times 10^{-11}\) GeV, where the uncertainty is the quadratic sum of contributions from \(\Delta\alpha_s(m_Z) = \pm 0.0009\), \(\Delta\Gamma_{J/\psi\to e^+e^-} = \pm 0.10\ \text{GeV}\), \(\Delta\overline{m}_c(\overline{m}_c) = \pm 0.0046\ \text{GeV}\), and the estimated magnitude of N\(^3\)LO contributions from Bayesian analysis. This work demonstrates for the first time how the PMC can be applied to obtain fixed-order perturbative predictions that are invariant under both renormalization and factorization scale variations.

\end{abstract}

\maketitle

As the only fundamental scalar particle in the Standard Model (SM) of particle physics, the Higgs boson was discovered by the ATLAS and CMS Collaborations at the LHC~\cite{ATLAS:2012yve, CMS:2012qbp}. Since then, increasingly precise measurements of its properties have established its central role in electroweak symmetry breaking and the generation of weak gauge boson masses, as summarized in Refs.\cite{ATLAS:2022vkf, CMS:2022dwd}.

Within the SM, fermion masses originate from Yukawa interactions with the Higgs field. To date, experimental measurements have firmly established the Higgs couplings to third-generation charged fermions, with results in excellent agreement with SM predictions. In contrast, the Higgs Yukawa couplings to first- and second-generation quarks remain only weakly constrained experimentally. A major challenge is that probing Higgs-quark couplings via the inclusive decay \(H\to q\bar{q}\) suffers from both overwhelming multijet backgrounds and strong quark mass suppression. As pointed out in Refs.\cite{Bodwin:2013gca, Kagan:2014ila}, the rare exclusive decay \(H\to J/\psi+\gamma\), despite a much smaller branching fraction than the inclusive mode, exhibits distinctive experimental signatures that greatly suppress QCD backgrounds~\cite{ATLAS:2022rej}. This renders it a promising probe of the Higgs Yukawa couplings to light and intermediate-mass quarks.

The decay amplitude for \(H\to J/\psi+\gamma\) admits two distinct mechanisms~\cite{Keung:1983ac, Jia:2008ep, Bodwin:2013gca}: a direct contribution mediated by the Higgs–quark Yukawa coupling, and an indirect one arising at one loop in the SM via \(H\to\gamma\gamma^{*}\), followed by fragmentation of the virtual photon into the vector quarkonium state. The indirect contribution dominates the decay rate; a leading-order analysis including quantum interference between the two amplitudes was presented in Ref.\cite{Bodwin:2013gca}. In this Letter, we focus exclusively on obtaining precise next-to-next-to-leading order (N\(^2\)LO) QCD predictions for the direct production mechanism, establishing \(H\to J/\psi+\gamma\) as a promising channel to probe the $Hc\bar{c}$ coupling at the LHC or other high-luminosity Higgs factories.

Early theoretical work on the \(J/\psi\) direct production mechanism originated with Shifman and Vysotsky in 1980~\cite{Shifman:1980dk}. The leading-order (LO) contribution to \(H\to J/\psi+\gamma\) was later computed in the non-relativistic quantum chromodynamics (NRQCD) framework~\cite{Jia:2008ep}. At each perturbative order, the corresponding short-distance coefficients (SDCs) exhibit large collinear logarithms \(\alpha_{s}^{n}\ln^{n}(m_{H}^{2}/m_{c}^{2})\) and divergent renormalon terms \(n!\beta_0^n\alpha_s^n\), which degrade the convergence of the perturbative series and require resummation for reliable predictions. Over the past decade, systematic refinements have extended these calculations to higher orders in the heavy-quark velocity expansion, including \({\cal O}(v^{2})\) and \({\cal O}(v^{4})\) corrections~\cite{Bodwin:2014bpa, Brambilla:2019fmu}. Combining NRQCD factorization with light-cone factorization -- along with the resummation of collinear logarithms~\cite{Konig:2015qat, Bodwin:2016edd} -- has substantially improved the precision of theoretical predictions. Additional large logarithms from Yukawa coupling renormalization are systematically resummed within the running \(\overline{\rm MS}\) mass scheme~\cite{Braaten:1980yq, Drees:1990dq, Djouadi:1997rp}.

At present, the SDCs for the total decay width \(\Gamma(H\to J/\psi+\gamma)\) in the direct production mechanism have been calculated up to next-to-leading order (NLO)~\cite{Zhou:2016sot} and next-to-next-to-leading order (N\(^2\)LO)~\cite{Jia:2024ini} in perturbative QCD.  It has been found that the N\(^2\)LO prediction still shows a strong dependence on the renormalization scale (\(\mu_{r}\)). In the conventional method, \(\mu_r\) is set to a characteristic momentum follow (\(Q\)) of the process in order to suppress large logarithmic terms in the perturbative coefficients and improve convergence, and is then varied over an interval such as \([Q/\eta,\eta Q]\) (with \(\eta = 2,3,4,...\)) to estimate theoretical uncertainties. However, this procedure is inherently {\it ad hoc} and introduces an artificial mismatch between the perturbative expansion coefficients and the running strong coupling, leading to an unphysical renormalization-scale dependence in the pQCD result. In fact, the conventional scale-setting method violates renormalization group invariance and thus limits the precision and reliability of fixed-order perturbative predictions.

The principle of maximum conformality (PMC)~\cite{Brodsky:2011ta, Brodsky:2012rj, Brodsky:2011ig, Mojaza:2012mf, Brodsky:2013vpa} provides a systematic, process-independent framework for removing the conventional ambiguities associated with renormalization scale setting and for exposing the intrinsic conformal structure of pQCD series. It extends the Brodsky-Lepage-Mackenzie method~\cite{Brodsky:1982gc} for scale-setting in pQCD to all orders, and it reduces analytically to the standard scale-setting procedure of Gell-Mann and Low~\cite{Gell-Mann:1954yli} in the QED Abelian limit~\cite{Brodsky:1997jk}. Specifically, the running behavior of the strong coupling constant is governed by the renormalization group equation (RGE), and the \(\{\beta_i\}\)-terms in the pQCD series encode the nonconformal contributions arising from renormalization. By exploiting this structure, the PMC unambiguously identifies and absorbs all RGE-related \(\{\beta_i\}\)-dependent terms into the running coupling at each order, thereby fixing the effective scale and determining the proper value of \(\alpha_s\) at every order. 

It is important to stress that not all the \(n_f\)-dependent terms should be identified with \(\{\beta_i\}\)-terms~\cite{Yan:2023hra}. A well-known example is provided by the so-called ``light-by-light"-type contributions, which arise from ultraviolet-finite diagrams and are thus unrelated to the renormalization of the strong coupling. If additional \(n_f\)-dependent terms arise from other scale-dependent quantities, such as long-distance matrix element (LDME), parton distribution functions, or running quark masses, these terms should also be treated as ``conformal" in the determination of \(\alpha_s\). The \(n_f\)-terms associated with the renormalization of the charm-quark Yukawa coupling provide another such example. Instead, they should be appropriately absorbed to fix the magnitudes of these parameters via their own RGEs or anomalous dimensions~\cite{Yan:2024oyb}, which can be implemented in a step-by-step manner. As a consequence, the PMC-improved perturbative coefficients are explicitly independent of the initial renormalization-scale choice and exhibit conformal behavior consistent with a theory possessing a vanishing \(\beta\)-function. Furthermore, via commensurate scale relations~\cite{Brodsky:1994eh, Huang:2020gic}, the PMC predictions are invariant under changes of renormalization scheme~\cite{Wu:2014iba, Wu:2015rga, Wu:2019mky}, yielding a self-consistent, scheme-independent fixed-order pQCD expansion with enhanced convergence and predictive power.

The total decay width for the direct production of \(J/\psi\) via the process $H\to J/\psi+\gamma$ can be expressed as
\begin{align}\label{gammatot}
	\Gamma_{\rm dir} =\frac{(m_{H}^2-m_{J/\psi}^2)^3}{8\pi m_{H}^3} \left|F_{\rm dir}\right|^2 
\end{align}
where $m_{H}$ and $m_{J/\psi}$ denote the masses of the Higgs boson and the $J/\psi$ meson, respectively. Within the NRQCD factorization framework~\cite{Bodwin:1994jh}, the form factor $F_{\rm dir}$ can be factorized into SDCs and LDMEs. This yields the total decay width from the direct production mechanism at N\(^2\)LO accuracy as
\begin{widetext}
\begin{eqnarray}
\label{gammaconv}
    \left|F_{\rm dir}\right|^2 &=& \bigg(\frac{g_{\rm Y}}{\sqrt{2}}\frac{4ee_{c}}{m_{H}^2} \frac{1}{1-\tau}\bigg)^2  
    \left( 1 + f_{1}(\mu_{r},\mu_{\Lambda}) \alpha_{s}(\mu_{r}) + f_{2}(\mu_{r},\mu_{\Lambda}) \alpha^{2}_{s}(\mu_{r}) \right) \times \left| \frac{\langle J/\psi({\bm \epsilon}) |\psi^{\dagger}{\bm \sigma}\cdot{\bm \epsilon}\chi(\mu_{\Lambda})|0\rangle} {\sqrt{2 m_c}} \right|^{2},
\end{eqnarray}
\end{widetext}
where $\mu_{r}$ and $\mu_{\Lambda}$ denote the renormalization and the factorization scales, respectively. The charm-quark Yukawa coupling is $g_{\rm Y} = (\sqrt{2}G_F)^{1/2}m_{c}$, with $m_{c}$ the charm-quark pole mass. Here $e_{c}=+2/3$ is the electric charge of the charm quark in units of the positron charge, and $\tau=4 m_{c}^2/m_{H}^2$. The perturbative coefficients $f_{1}$ and $f_{2}$ can be found in Refs.\cite{Zhou:2016sot, Jia:2024ini}. 

First, it has been noted that some of the large logarithms \(\ln\left( m_{H}^{2}/m_{c}^{2} \right)\) in the expansion coefficients \(f_{1,2}\) are closely associated with the renormalization of the charm-quark Yukawa coupling; these can be resummed to all orders to obtain a precise value for the charm-quark Yukawa coupling. Since \(g_{\rm Y} \propto m_{c}\), this resummation is equivalent to converting the charm-quark pole mass \(m_{c}\) appearing in the Yukawa coupling \(g_{\rm Y}\) to the \(\overline{\rm MS}\)-scheme running charm-quark mass evaluated at the Higgs mass scale \(\overline{m}_c(m_{H})\)~\cite{Braaten:1980yq, Drees:1990dq, Djouadi:1997rp}, which can be demonstrated using the perturbative relation between the pole mass and the \(\overline{\rm MS}\)-scheme running quark mass, combined with the \(\overline{\rm MS}\)-scheme quark mass anomalous dimension.

Second, the factorization-scale-dependent color-singlet LDME $\langle J/\psi({\bm \epsilon})|\psi^{\dagger}{\bm \sigma}\cdot{\bm \epsilon}\chi(\mu_{\Lambda})|0\rangle$ is non-perturbative yet universal. A model-independent approach to determining such LDME is to extract it from experimental observables combined with perturbatively calculable quantities~\cite{Huang:2023pmn}. Specifically, this LDME can be extracted from the \(J/\psi\) leptonic decay width via the following relation:
\begin{widetext}
\begin{eqnarray}
\label{Gammall}
    \Gamma_{J/\psi \to e^{+}e^{-}} &=& \frac{8\pi\alpha^{2}(m_{J/\psi})e_{c}^{2}}{3m_{J/\psi}^{2}} \left|{\cal C}\left(\mu_{r},\mu_{\Lambda}\right)\right|^{2}  \times|\langle J/\psi({\bm \epsilon})|\psi^{\dagger}{\bm \sigma}\cdot{\bm \epsilon}\chi(\mu_{\Lambda})|0\rangle|^{2},
\end{eqnarray}
\end{widetext}
where the SDCs ${\cal C}\left(\mu_{r},\mu_{\Lambda}\right)$ are known up to N\(^3\)LO accuracy~\cite{Beneke:1997jm, Czarnecki:1997vz, Kniehl:2006qw, Marquard:2006qi, Marquard:2009bj, Marquard:2014pea, Feng:2022vvk}. Then the N\(^2\)LO-level total decay width (\ref{gammatot}) can be reformulated as
\begin{align}\label{Gconv}
    \Gamma_{\rm dir}(H\to J/\psi+\gamma) = C_{0} \overline{m}_{c}^{2}(m_{H})R(\mu_{r}),
\end{align}
where the overall factor
\begin{displaymath}
C_{0}=\frac{3 G_F \alpha(m_H/2) m_{J/\psi}^2(m_H^2-m_{J/\psi}^2)^3}{2\sqrt{2} \pi\alpha^2(m_{J/\psi}) m_c m_H^7(1-\tau)^2} \Gamma_{J/\psi\to e^+e^-}.
\end{displaymath}
The LDME determined from Eq.(\ref{Gammall}) evolves with the factorization scale according to its own RGE. By inverting the corresponding \(\{\beta_i\}\)-terms in the perturbative series directly associated with the RGE of the LDME, we can unambiguously extract the physical, factorization-scale-independent component of the series. This conclusion follows from the fact that the leading-order anomalous dimensions of the relevant NRQCD vector current are identical in both cases~\cite{Feng:2022vvk, Jia:2024ini}, first appearing at N\(^2\)LO. This explains the absence of factorization-scale dependence on the left-hand side of Eq.(\ref{Gconv}), and thus eliminates a dominant source of theoretical uncertainty present in conventional treatments based on potential-model wave functions or arising from the tacit use of a unified LDME regardless of the factorization scale. Furthermore, the renormalization-scale dependent factor takes the form:
\begin{align}\label{Rconv}
    R(\mu_{r}) = \sum_{i=0}^{2}r_{i}(\mu_{r})\alpha_{s}^{i}(\mu_{r}),
\end{align}
where \(r_{0}=1\), the NLO and N\(^2\)LO coefficients \(r_{1,2}(\mu_{r})\) (with \(n_f\) the number of active flavors) are
\begin{eqnarray}
r_{1} &=& 0.8861, \\
r_{2}(\mu_r) &=& \left(0.5512+0.0431\ln\frac{\mu_r^2}{m_H^2}+0.0901\ln\frac{\mu_r^2}{m_c^2}\right) n_f \nonumber \\
 	  & &-12.8116-0.7104\ln\frac{\mu_r^2}{m_H^2} +1.4860\ln\frac{\mu_r^2}{m_c^2}.
\end{eqnarray}

For numerical calculations, the following values are adopted as the input parameters~\cite{Jia:2024ini}
 \begin{align}
	m_{H}&=125.20~{\rm GeV},  &m_c&=1.50~{\rm GeV}, \notag\\
	m_{J/\psi}&=3.0969~{\rm GeV},   &\overline{m}_c(\overline{m}_c)&=1.2730~{\rm GeV}, \nonumber
\end{align}
and $\alpha_s(m_Z)=0.1180\pm0.0009$. The \(J/\psi\) leptonic decay width is taken as $\Gamma_{J/\psi \to e^{+}e^{-}} = 5.53 \pm 0.10$~keV~\cite{ParticleDataGroup:2024cfk}. We use the Fermi constant $G_F=1.1664\times 10^{-5}~{\rm GeV}^{-2}$, and the QED running coupling at two scales: $\alpha(m_{H}/2)=1/128.46$ and $\alpha(m_{J/\psi})=1/132.77$.

\begin{figure}[htb] 
	\centering
	\includegraphics[width=0.48\textwidth]{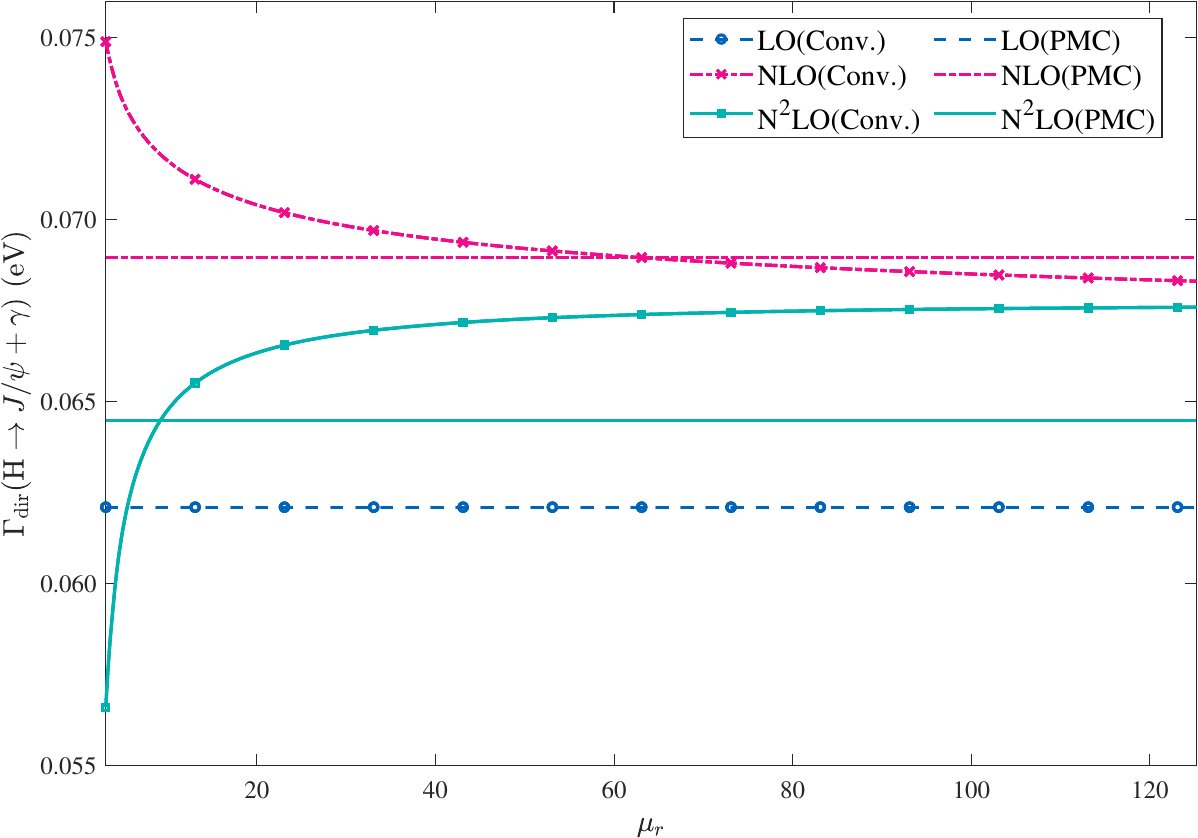}
	\caption{The $\mu_{\Lambda}$-independent total decay width $\Gamma_{\rm dir}(H\to J/\psi+\gamma)$ up to N\(^2\)LO QCD corrections versus the renormalization scale $\mu_r$ under the conventional and PMC scale-setting methods, respectively. The LO results are the same. The PMC fixed-order predictions are scale-invariant.}
	\label{pgammaconv}
\end{figure}

We present the $\mu_{\Lambda}$-independent total decay width up to N\(^2\)LO QCD corrections versus the renormalization scale \(\mu_r\) in the conventional scale-setting method [see Eq.\eqref{Gconv}] in \autoref{pgammaconv}, where the PMC predictions are given as a comparison. As shown, the previously unphysical negative total decay width is removed and the uncertainties associated with the factorization scale are eliminated, both of which highlight the importance of separately treating the RGE terms for the charm-quark Yukawa coupling and the LDME, respectively. Nevertheless, the resulting prediction still exhibits pronounced renormalization-scale dependence at N\(^2\)LO accuracy -- especially in the small-\(\mu_r\) region. When taking \(\mu_r \in [m_{J/\psi},m_{H}]\), the scale error for the total decay width under the conventional scale-setting method is approximately \(16.31\%\), e.g.
\begin{align} \label{gammacv}
	\Gamma_{\rm dir}(H\to J/\psi+\gamma)|_{\rm Conv.} = 6.7500^{+0.0021}_{-1.0807}\times 10^{-11}~{\rm GeV}.
\end{align}

Following the standard PMC single-scale-setting procedures~\cite{Shen:2017pdu, Yan:2022foz}, one can further deal with the RGE-terms associated with the strong coupling with the purpose of eliminating the renormalization scale dependence in Eq.(\ref{Gconv}). For the purpose, we can transform the initial $n_f$-series (\ref{Rconv}) into the $\{\beta_{i}\}$-series by using the general degeneracy relations among different orders~\cite{Bi:2015wea}, which yields the expansion coefficients:
\begin{equation}
    r_{1} = r_{1,0}, \;\; r_{2}(\mu_{r}) = r_{2,0} + r_{2,1}(\mu_{r}) \beta_{0}.
\end{equation}
Here, $\beta_{0} = (11-2n_{f}/3)/(4\pi)$ is the first coefficient of QCD $\beta$-function, $r_{i,0}$ are scale-invariant conformal coefficients, and $r_{2,1} = \hat{r}_{1,0} + \hat{r}_{2,1} \ln(\mu_{r}^{2}/Q^{2})$ is scale-dependent non-conformal coefficient with $\hat{r}_{i,j} = r_{i,j}(\mu_{r}=Q)$, specially $\hat{r}_{i,0} = r_{i,0}$. The RGE-induced non-conformal term \(r_{2,1}(\mu_{r}) \beta_{0}\) can then be used to determine an overall effective strong coupling $\alpha_{s}(Q_{*})$, where $Q_{*}$ denotes the PMC scale and corresponds to the effective momentum flow for the process. The resulting scale-invariant PMC series is thus given by
\begin{align}\label{Rpmc}
    R(\mu_r)|_{\rm PMC} = 1 + \sum_{i=1}^{2} r_{i,0} \alpha_{s}^{i}(Q_{*}),
\end{align}
where $Q_{*}$ is given by
\begin{displaymath}
    \ln \frac{Q_{*}^{2}}{Q^{2}} = -\frac{\hat{r}_{2,1}}{\hat{r}_{1,0}} \alpha_{s}(Q_{*}),
\end{displaymath}
which is determined at the leading-log (LL) accuracy. Setting all the inputs to their central values except for scale variation, we obtain \(Q_*^{\rm LL} = 9.1706~{\rm GeV}\), yielding \(R(\mu_r)|_{\rm PMC} = 1.0382\). Such scale-invariance is confirmed by \autoref{pgammaconv}. This demonstrates that, if all the RGE-related non-conformal \(\{\beta_i\}\) terms in the perturbative series are properly absorbed into \(\alpha_s(Q_*)\), the resulting series~\eqref{Rpmc} will further become scheme- and scale-independent~\cite{Wu:2019mky}. We then obtain
\begin{align}
	\Gamma_{\rm dir}(H\to J/\psi+\gamma)|_{\rm PMC} = 6.4574\times10^{-11}~{\rm GeV}. \label{gpmc}
\end{align}

Furthermore, the elimination of divergent renormalon terms leads to a significantly improved convergence of the PMC conformal series relative to the conventional scale-setting method~\cite{Wu:2013ei, Wu:2018cmb, DiGiustino:2023jiq}. To further characterize the convergence behavior, we define a $\kappa$-factor for the N$^2$LO series \eqref{Rconv} or \eqref{Rpmc}, that is
\begin{align}
	\kappa_{i}|_{\rm Conv.} &=  \left| \left(r_{i}(\mu_r)/r_{0}\right) \alpha_{s}^{i}(\mu_r) \right|, \label{kappaconv} \\
	\kappa_{i}|_{\rm PMC} &= \left| \left(\hat{r}_{i,0}/\hat{r}_{0,0}\right) \alpha_s^{i}(Q_*)\right|. \label{kappapmc}
\end{align}
These ratios (\(\kappa=\{\kappa_i\}\)) of contributions from different orders to the LO contribution characterize the convergence of the perturbative series. Numerically, we have
\begin{align}
	\kappa|_{\rm PMC} &= \{1,0.1613, 0.1231\},\quad  \forall \mu_r \label{kappa1}  \\ 
	\kappa|_{\rm Conv.} &= \{1,0.2240,0.3126\},\quad \mu_r = m_{J/\psi}   \label{kappa2}   \\
	\kappa|_{\rm Conv.} &= \{1,0.1109,0.0257\},\quad \mu_r = m_{H}/2      \label{kappa3}\\ 
	\kappa|_{\rm Conv.} &= \{1,0.0998,0.0113\}.\quad \mu_r = m_{H}     \label{kappa4}
\end{align}
The convergence of the conventional series is highly scale-dependent and strongly tied to the magnitude of \(\alpha_s\); it converges well for an appropriate choice of \(\mu_r\), yet diverges at low scales where \(\alpha_s\)-power suppression cannot overcome the renormalon-induced growth of expansion coefficients. In contrast, the convergence of the PMC series is scale-independent -- a property that reflects the intrinsic nature of perturbative series.

Furthermore, for any truncated perturbative series, intrinsic theoretical uncertainties inevitably arise from the unknown higher-order (UHO) contributions. Since the exact all-order pQCD result is not available, it is natural to characterize the impact of these UHO terms in a probabilistic manner. When applying Bayesian analysis (BA) to the pQCD series~\cite{Cacciari:2011ze, Bagnaschi:2014wea, Bonvini:2020xeo}, a prior probability distribution is first assigned to the unknown perturbative coefficients; this distribution is then systematically updated via Bayes' theorem as additional perturbative information is incorporated. It has been shown that the PMC-improved perturbative series -- which is typically more convergent, renormalization-scheme invariant, and renormalization-scale invariant -- provides a significantly more robust and reliable basis than conventional series for estimating the magnitude of UHO contributions~\cite{Shen:2022nyr, Shen:2023qgz, Luo:2023cpa, Yan:2023mjj, Yan:2024hbz, Li:2025hyh}. In this work, we therefore use the Bayesian approach to quantify the uncertainty arising from the unknown N\(^3\)LO contributions within the PMC-improved perturbative series. A detailed explanation of the combined use of the PMC and the BA method can be found in Refs.\cite{Yan:2022foz, Duhr:2021mfd, Shen:2022nyr, Shen:2023qgz}. More explicitly, applying the BA method with a fixed degree-of-belief (DoB=95.5\%) for the total decay width and using the known coefficients up to a fixed order \{$r_1,r_2,r_3...$\}, one can obtain the estimated UHO-terms coefficients $r_{p+1}$ within a specific credible interval (CI) of $r_{p+1}\in [-r_{p+1}^{\rm (DoB)},r_{p+1}^{\rm (DoB)}]$, where
\begin{align}
	r_{p+1}^{(\rm DoB)} = \left\{
    \begin{aligned}
		&\bar{r}_{(p)}\frac{p+1}{p}{\rm DoB}, & {\rm DoB} \le \frac{p}{p+1}\\
		&\bar{r}_{(p)}[(p+1)(1-{\rm DoB})]^{-1/p}, & {\rm DoB} 
		\geq \frac{p}{p+1}
	\end{aligned} \right.
\end{align}
with $\bar{r}_{(p)}=\rm max\{|r_1|,|r_2|,|r_3|...\}$. In particular, the next-order prediction for the physical quantity $\rho_{p+1}$ for the general pQCD series truncated at the $p_{th}$-order, $\rho_p=\sum_{i=1}^{p} r_i\alpha_s^i$ satisfies the following containment relation:
\begin{displaymath}
	\rho_{p+1}\in[\rho_{p}-r_{p+1}^{(\rm DoB)}\alpha_s^{p+1},\rho_p+r_{p+1}^{(\rm DoB)}\alpha_s^{p+1}].
\end{displaymath}

\begin{figure} [htb]
\centering
\includegraphics[width=0.48\textwidth]{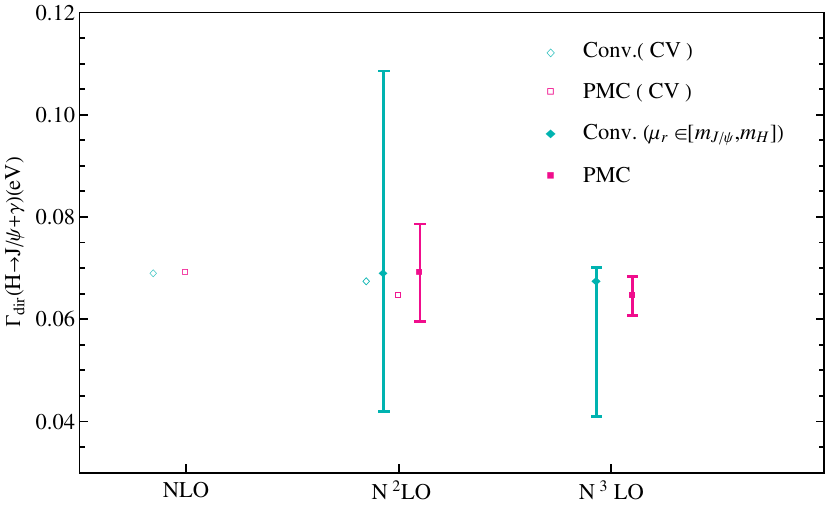}
\caption{Comparison of the calculated central values (CV) of the total decay width \(\Gamma_{\rm dir}(H\to J/\psi+\gamma)\) with the predicted credible intervals of the corresponding pQCD approximants (with DoB = \(95.5\%\)) up to N\(^3\)LO. The green hollow diamonds and red hollow squares represent the central values of the known fixed-order pQCD predictions obtained using the conventional (Conv.) and PMC scale-setting methods, respectively. The green solid diamonds and red solid squares with error bars represent the predicted credible intervals obtained using the BA method.}
	\label{uho}
\end{figure} 

We present a comparison of the calculated central values of the total decay width \(\Gamma_{\rm dir}(H\to J/\psi+\gamma)\) with the predicted CIs of the corresponding pQCD approximants (with DoB = \(95.5\%\)) up to N\(^3\)LO in \autoref{uho}. The CIs are predicted using the BA method, which is applicable when at least the first two terms of the perturbative series are known; accordingly, the predictions correspond to the N\(^2\)LO and N\(^3\)LO terms. For the scale-dependent conventional series, the scale uncertainty is estimated using $\mu_r\in[m_{J/\psi},m_{H}]$. \autoref{uho} shows that the probability distribution becomes more accurate and the resulting CIs shrink for the same DoB as more loop terms are included, implying that the predicted CIs for N\(^2\)LO and N\(^3\)LO exhibit a clear convergent trend. Note that, owing to the divergent behavior of the conventional pQCD series at low scales, the PMC series yields more precise predictions for the UHO terms via the BA method. More explicitly, using the BA method to estimate the UHO terms, the predicted N\(^3\)LO UHO-contributions are
\begin{align}
	\Delta\Gamma_{\rm dir}|_{\rm UHO}^{\rm Conv.} &=\left(_{-2.6489}^{+0.2578}\right)\times10^{-11}~{\rm GeV},\\
	\Delta\Gamma_{\rm dir}|_{\rm UHO}^{\rm PMC} &=\left(_{-0.3793}^{+0.3793}\right)\times10^{-11}~{\rm GeV}.
\end{align}

In addition to the uncertainties from the renormalization scale and the possible contribution from the UHO terms, there remain several sources of uncertainty, such as $\Delta\alpha_s(m_Z), \Delta\overline{m}_c(\overline{m}_c)$, and the uncertainty in the measured \(J/\psi\) leptonic decay width $\Delta\Gamma_{J/\psi\to e^+e^-}$. For convenience, when evaluating any one of these uncertainties, all other input parameters are set to their central values.

The uncertainties for the two overall parameters \(\Delta\overline{m}_c(\overline{m}_c)=\pm0.0046~{\rm GeV}\) (denoted as \(\Delta\overline{m}_c\)) and \(\Delta\Gamma_{J/\psi\to e^+e^-}=\pm0.10~{\rm keV}\) (denoted as \(\Delta\Gamma_{J/\psi}\)) are
\begin{align}
	\Delta\Gamma_{\rm dir}|^{\rm Conv.}_{\Delta\overline{m}_c,\rm \Delta\Gamma_{J/\psi}} =&\, (^{+0.0487+0.1217}_{-0.0487-0.1217})\times10^{-11} ~{\rm GeV}, \\
	\Delta\Gamma_{\rm dir}|^{{\rm PMC}}_{\Delta\overline{m}_c,\rm \Delta\Gamma_{J/\psi}} =&\, (^{+0.0466+0.1164}_{-0.0466-0.1164})\times10^{-11} ~{\rm GeV}.
\end{align}
The uncertainties arising from \(\Delta\alpha_s(m_Z)=\pm0.0009\) are
\begin{align}
	\Delta\Gamma_{\rm dir}|^{\rm Conv.}_{\Delta\alpha_s(m_Z)} =&\, (^{+0.0030}_{-0.0030})\times10^{-11}~{\rm GeV}  \label{aconv}\\ 
	\Delta\Gamma_{\rm dir}|^{\rm PMC}_{\Delta\alpha_s(m_Z)} =&\, (^{+0.0063}_{-0.0066})\times10^{-11}~{\rm GeV}  \label{apmc}
\end{align}
Eqs.(\ref{aconv},\ref{apmc}) show the PMC prediction is more sensitive to the value of \(\Delta\alpha_s(m_Z)\). This is reasonable, since the PMC uses the process-dependent RGE terms to set the effective value of \(\alpha_s\); consequently, the resulting pQCD series is more sensitive to the precise value of \(\alpha_s\). 

Including all these sources of uncertainty, our final predictions for the N\(^2\)LO total decay width for the direct production of \(J/\psi\) in the process $H\to J/\psi+\gamma$ are
\begin{align}
\Gamma_{\rm dir}|_{\rm tot}^{\rm Conv.} =&\, (6.7500^{+0.2893}_{-2.6522})\times10^{-11}~{\rm GeV}  \label{errconv}\\ 
\Gamma_{\rm dir}|_{\rm tot}^{\rm PMC} =&\, (6.4574^{+0.3995}_{-0.3995})\times10^{-11}~{\rm GeV}  \label{errpmc}
\end{align}
where the net uncertainties are squared averages of those from $\mu_r\in[m_{J/\psi},m_H],\Delta\alpha_s(m_Z),\Delta\overline{m}_c(\overline{m}_c),\Delta\Gamma_{J/\psi}$, and the predicted N$^3$LO-terms.

In summary, we have applied the PMC scale-setting method to the total decay width for \(H\to J/\psi+\gamma\) via the \(J/\psi\) direct-production mechanism, at N$^2$LO accuracy. Following the PMC principle, different types of RGEs should be treated separately to determine the correct effective values of the corresponding running parameters, such as the charm-quark Yukawa coupling, the LDME, and the strong coupling. With these treatments, one obtains a consistent higher-order pQCD prediction free of both renormalization- and factorization-scale ambiguities. Such a high-precision prediction firmly establishes \(H\to J/\psi+\gamma\) as a promising channel for probing the $Hc\bar{c}$ coupling at the LHC and other high-luminosity Higgs factories. Furthermore, the present approach demonstrates for the first time that the PMC strategy -- using the RGEs to determine the correct effective values of scale-dependent running parameter -- also provides a mechanism to eliminate the standard factorization-scale dependence. The PMC procedure thus improves the precision of SM tests and enhances the sensitivity to new phenomena. The PMC method can be applied to a wide variety of perturbatively calculable collider and other processes.

\hspace{1cm}

\noindent{\bf Acknowledgements}: This work was supported in part by the Natural Science Foundation of China under Grants No.12547115, No.12575080 and No.12547101, and by the Chongqing Natural Science Foundation under Grant No. CSTB2025NSCQ-GPX0745.

\end{document}